\documentclass[twocolumn,aps,prc,reprint]{revtex4-1}
\usepackage{blindtext}
\usepackage{graphicx}
\usepackage{dcolumn}
\usepackage{bm}
\usepackage{booktabs}
\usepackage{color}
\usepackage{mathrsfs}
\usepackage{ulem}
\usepackage{verbatim}
\usepackage{indentfirst}
\usepackage{float}
\usepackage[colorlinks,
           bookmarks=true,
           linkcolor=blue,
           urlcolor=blue,
            anchorcolor=black,
            citecolor=blue
            ]{hyperref}
\usepackage[bf,raggedright,indentafter]{titlesec}
\usepackage{hypcap}

\usepackage{epsfig}
\graphicspath{{plots/}}

\begin{document}
\title{Survival rate of initial azimuthal anisotropy in a multi-phase transport model}

\author{Liang Zhang$^1$}
\email[Electronic address:~]{l.zhang@mails.ccnu.edu.cn}
\author{Feng Liu$^1$}
\email[Electronic address:~]{fliu@mail.ccnu.edu.cn}
\author{Fuqiang Wang$^{1,2}$}
\email[Electronic address:~]{fqwang@purdue.edu}

\address{%
$^1$ Key Laboratory of Quark and Lepton Physics (MOE) and Institute of Particle Physics,\\
Central China Normal University, Wuhan 430079, China\\
$^2$Department of Physics and Astronomy, Purdue University, West Lafayette, Indiana 47907, USA
}
\begin{abstract}
We investigate the survival rate of an initial momentum anisotropy (${v}_2^{ini}$), not spatial anisotropy, to the final state in a multi-phase transport (AMPT) model in Au+Au collisions at $\sqrt{s_{NN}}$=200~GeV. It is found that both the final-state parton and charged hadron $v_2$ show a linear dependence versus $v_2^{ini}\{\rm PP\}$ with respect to the participant plane (PP). It is found that the slope of this linear dependence (referred to as the survive rate) increases with transverse momentum ($p_T$), reaching~$\sim$100\% at $p_T$$\sim$2.5 GeV/c for both parton and charged hadron. The survival rate decreases with collision centrality and energy, indicating decreasing survival rate with increasing interactions. It is further found that a $v_2^{ini}\{\rm Rnd\}$ with respect to a random direction does not survive in $v_2\{\rm PP\}$ but in the two-particle cumulant $v_2\{2\}$. The dependence of $v_2\{2\}$ on $v_2^{ini}\{\rm Rnd\}$ is quadratic rather than linear.
\end{abstract}
\maketitle

\section{Introduction}
A new state of matter, the strongly coupled quark gluon plasma (sQGP), is created in relativistic heavy ion collisions~\cite{Adams:2005dq,adcox2005formation,arsene2005quark,back2005phobos}. 
One of the most important evidence is the measured large elliptic flow in non-central heavy ion collisions, believed to stem out of final state interactions in the anisotropic overlap zone~\cite{Ollitrault:1992aa}. 
The measured elliptic flow is so large that it is compatiable with hydrodynamic calculations with minimal shear viscosity to entropy density ratio ($\eta/s$), indicating maximal final-state interactions~\cite{PRL2004DE,Heinz:2013th}.

Present hydrodynamic calculations start from an initial condition of isotropic momentum distribution. It has been argued, however, that the initial momentum anisotropy may not be zero in relativistic heavy ion collisions. 
For example, it is suggested that the wave function is asymmetric in momentum space due to Heisenberg uncertainty principle because of the spatial anisotropic overlap~\cite{fqw2014mom}.
In classical Yang-Mills dynamics it is found that initial momentum anisotropy can arise from the event-by-event breaking of rotational invariance in local domains whose size is determined by the saturation scale~\cite{Schenke:2015aqa}.
Initial flow in classical Yang-Mills field can also develop from the non-abelian generalization of Gauss' Law and Ampere's and Faraday's Laws~\cite{Fries:2014oya}.
In proton-proton collisions color reconnection can produce initial flow-like correlations~\cite{Ortiz:2013yxa} and it may be relevant for heavy ion collisions as well. 
If there indeed exist initial momentum anisotropies and these initial anisotropies can partially survive to the final state, then the comparison of data to hydrodynamics without initial momentum anisotropy would not be reliable to extract transport properties of the sQGP, such as the $\eta/s$.
In this paper, we investigate the survival rate of an input initial momentum anisotropy using a parton transport model.

\section{Analysis}
We employ A Multi-Phase Transport (AMPT) model with string melting and 3 mb parton cross section~\cite{lin2002partonic,lin2005multiphase}.  
This model can describe well the measured particle rapidity distributions, transverse momentum spectra, and elliptic flow~\cite{Lin:2014tya}.
AMPT consists of four main parts:  the initial condition, parton-parton interactions, hadronization, and hadronic scatterings. The initial condition is taken from the HIJING model~\cite{Wang:1991hta}. It uses Glauber nuclear geometry to model the spatial and momentum information of minijet partons from hard processes and strings from soft processes. The interactions of partons are treated by the ZPC parton cascade model~\cite{zhang1998zpc}. After parton interactions cease, a combined coalescence and string fragmentation model is used for the hadronization of partons. Finally, the ART model is used to describe the elastic and inelastic scatterings of hadrons~\cite{Li:1995pra}.

Elliptic flow can be quantified by $v_2$, the second harmonic Fourier coefficient of the particle azimuthal distribution in momentum space~\cite{Voloshin:1994mz},
\begin{eqnarray}\label{eq:definition1}
dN/d\phi \propto 1+2v_2\{\text{PP}\}\cos2(\phi-\Psi_2\{\text{PP}\}).
\end{eqnarray} 
In AMPT, the initial parton azimuthal distribution is isotropic:
\begin{eqnarray}
\frac{dN}{d\phi_{ini}} =\text{constant}\text{.}
\end{eqnarray} 
We can artificially create a momentum anisotropy by ``squeezing'' particles towards a particular plane. We first choose this plane to be the participant plane (PP) of the initial partons in configuration space. The azimuthal angle of the participant plane is given by 
\begin{eqnarray}
\displaystyle \Psi_2^{\text{PP}}=\frac{\text{atan2}(\langle r_{ini}^2\sin 2\phi_{ini}\rangle,\langle r_{ini}^2\cos 2\phi_{ini})\rangle+\pi}{2}\text{,}
\end{eqnarray} 
where $r_{ini}$ and $\phi_{ini}$ are polar coordinate position. Mathematically we change each parton's initial azimuthal angle $\phi_{ini}$ into $\phi_{ini}'$ by:
\begin{eqnarray}
\phi_{ini}^\prime=\phi_{ini}+\delta\text{,}
\end{eqnarray} 
such that
\begin{eqnarray}
\frac{dN}{d\phi_{ini}^\prime}=1+2v_2^{ini}\{\text{\text{PP}}\}\cos2(\phi_{ini}^\prime-\Psi_2^{\text{PP}})\text{.}
\end{eqnarray} 
In order to achieve an initial anisotropy $v_2^{ini}\{\text{\text{PP}}\}$ with respect to $\Psi_2^{\text{PP}}$, one applies
\begin{eqnarray}\label{eq:5}
\delta=-\frac{v_2^{ini}\{\text{\text{PP}}\}\sin2(\phi_{ini}-\Psi_2^{\text{PP}})}{1+2v_2^{ini}\{\text{\text{PP}}\}\cos2(\phi_{ini}-\Psi_2^{\text{PP}})}\text{.}
\end{eqnarray} 

Second, we choose this particular plane to be a random azimuthal direction $\Psi_2^{\text{Rnd}}$, not the participant plane along which hydrodynamic collective flow develops. We denote this initial anisotropy as $v_2^{ini}\{\text{Rnd}\}$. Same as Eq.~\ref{eq:5}, on applies
\begin{eqnarray}\label{eq:6}
\delta=-\frac{v_2^{ini}\{\text{Rnd}\}\sin2(\phi_{ini}-\Psi_2^{\text{Rnd}})}{1+2v_2^{ini}\{\text{Rnd}\}\cos2(\phi_{ini}-\Psi_2^{\text{Rnd}})}\text{.}
\end{eqnarray} 

In these operations, only the parton's azimuthal angle is altered, no other changes. The event now has an initial anisotropy ($v_2^{ini}\{\text{\text{PP}}\}$ or $v_2^{ini}\{\text{Rnd}\}$). The event then evolves as modeled by AMPT. In this analysis we have used a given $v_2^{ini}$ in each event, independent of the parton $p_T$.

We analyze the momentum anisotropies of the final-state partons (i.e.~after parton interactions cease and before hadronization) and the final-state hadrons by the Fourier coefficients~\cite{poskanzer1998methods}:
\begin{eqnarray}
v_2\{\text{\text{PP}}\}=\langle \cos 2(\phi-\Psi_2^{\text{PP}})\rangle\text{,}
\end{eqnarray} 
where $\phi$ is the particle (parton or hadron) azimuthal angle.

Experimentally, however, the participant plane is inaccessible. The momentum anisotropy is often analyzed by final-state two-particle correlations. In absence of intrinsic particle correlations (nonflow), the final-state two-particle correlations are caused by each particle's correlation to the common participant plane (i.e.~flow correlations) of Eq.~\ref{eq:definition1}. In such a case, the two-particle density is given by
\begin{eqnarray}\label{eq:999}
d^2N/d\Delta\phi=1+2v_2\{2\}^2\cos\Delta\phi,
\end{eqnarray}
where $\Delta\phi=\phi_1-\phi_2$ is the azimuthal angle difference between the two particles.
The momentum anisotropy can thereby be calculated by
\begin{eqnarray}\label{eq:8}
v_2\{2\}=\sqrt{\langle\langle\cos2(\phi_1-\phi_2)\rangle\rangle},
\end{eqnarray}
without requiring the participant plane~\cite{Borghini:2000sa}. We use the Q-cumulant method~\cite{Bilandzic:2010jr} to calculate $\langle\cos2(\phi_1-\phi_2)\rangle$ in each event, by
\begin{eqnarray}
\langle\cos2(\phi_1-\phi_2)\rangle=\frac{|Q_2|^{2}-M}{M(M-1)}.
\end{eqnarray}
Here the second-harmonic flow vector $Q_2$ is given by 
\begin{eqnarray}\label{eq:qvector}
Q_2=\sum\limits_{i}^{M}e^{i2\phi_{i}},
\end{eqnarray}
and $M$ is the number of particles used in the Q-cumulant.\\
In the averaging of $\langle\cos2(\phi_1-\phi_2)$ over events in Eq.~\ref{eq:8}, we have applied the weight of $M(M-1)$. For differential flow as a function of $p_T$, we compute the correlation between a particle of interest at $p_T$(azimuth $\phi$) and a reference particle (azimuth $\phi_{\rm ref}$), $\langle\langle\cos2(\phi-\phi_{\rm ref})\rangle\rangle$. 
The anisotropy of the particles of interest is then given by
\begin{eqnarray}
v_2\{2\}(p_T)=\frac{\langle\cos2(\phi-\phi_{\rm ref})\rangle}{v_2^{\rm{ref}}\{2\}}
\end{eqnarray}
where the reference particle $v_2^{\rm{ref}}\{2\}$ is given by Eq.~\ref{eq:8}, where $\phi_1$ and $\phi_2$ are both of reference particles.
We again use the Q-cumulant method to compute 
$\langle\cos2(\phi-\phi_{\rm ref})\rangle$, by 
\begin{eqnarray}
\langle\cos2(\phi-\phi_{\rm ref})\rangle=\frac{p_2Q_2^*-m_q}{m_pM-m_q},
\end{eqnarray}
where $p_2$ is the second-harmonic flow vector (Eq.~\ref{eq:qvector}) of the particles of interest, $m_p$ is the number of the particles of interest, and $m_q$ is the number of particles in the overlap of the two groups (particles of interest and reference particles).

\section{Results}\label{sec:3}
\begin{figure}[!htb] 
  \begin{center}
    \includegraphics[height=5.5cm]{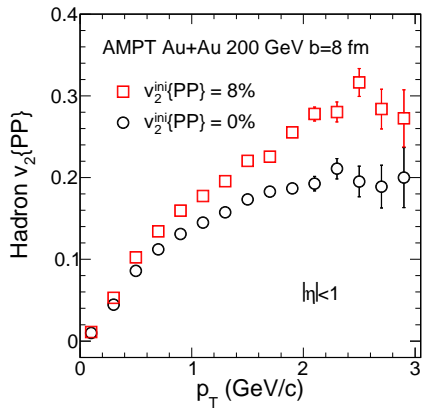}
    \caption{ (color online) Mid-rapidity ($|\eta| < 1$) hadrons $v_2\{\text{\text{PP}}\}$ versus $p_T$ with $v_2^{ini}\{\text{\text{PP}}\}=0\% $ and $v_2^{ini}\{\text{\text{PP}}\}=8\%$ for $b$=8~fm Au+Au collisions at $\sqrt{s_{NN}}$=200~GeV by AMPT (string melting).}
   \label{fig:1}
  \end{center}
\end{figure}

Figure~\ref{fig:1} shows the final-state charged hadron $v_{2} \{\text{\text{PP}}\}$ at mid-rapidity ($|\eta|<1$) versus transverse momentum ($p_T$) for fixed impact parameter ($b$=8~fm) Au+Au collisions at $\sqrt{s_{NN}}$=200~GeV.
The black circles show the default result without initial $v_2^{ini}\{\rm PP\}$. The red squares show the result with $v_2^{ini}\{\text{\text{PP}}\}$=8\%. With initial $v_2^{ini}\{\rm PP\}$, the final $v_2\{\rm PP\}$ is larger, and the effect is more significant at higher $p_T$. This is consistent with the finding in Ref.~\cite{he2015anisotropic} that higher $p_T$ partons suffer fewer collisions on average and thus retain larger fraction of their initial anisotropy.

\begin{figure}[!htb]
  \begin{center}
    \includegraphics[height=5.5cm]{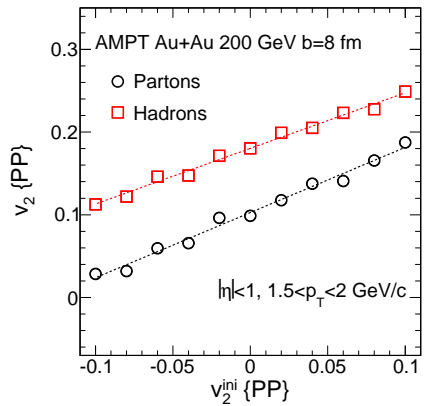}
    \caption{\label{fig:2}(color online) Mid-rapidity ($|\eta| < 1$) final-state partons and hadrons $v_2\{\text{\text{PP}}\}$ at $1.5<p_T<2$ GeV/c versus $v_2^{ini}\{\text{\text{PP}}\}$  for $b$=8~fm Au+Au collisions at $\sqrt{s_{NN}}$=200~GeV by AMPT (string melting).}
  \end{center}
\end{figure}

\begin{figure}[!htb]
  \begin{center}
    \includegraphics[height=5.5cm]{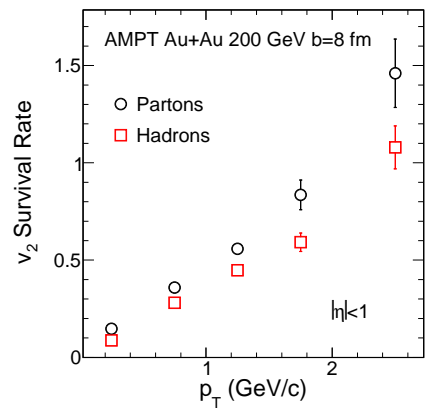}
    \caption{\label{fig:3}(color online) The survival rate of initial momentum anisotropy to final-state partons and charged hadrons at mid-rapidity ($|\eta<1|$) as a function of $p_T$ for $b$=8~fm Au+Au collisions at $\sqrt{s_{NN}}$=200~GeV by AMPT (string melting).}
  \end{center}
\end{figure}

Figure~\ref{fig:2} shows the parton and hadron $v_2\{\text{\text{PP}}\}$ versus initial parton $v_2^{ini}\{\text{\text{PP}}\}$, for 1.5$<p_T<$2 GeV/c as an example. We fit the results with the two-parameter linear function, $v_2\{\text{\text{PP}}\}=r\times v_{2}^{ini}\{\text{\text{PP}}\}+v_2(0)$. The fitting parameter $v_2(0)$ corresponds  to the result without an initial $v_2^{ini}\{\rm PP\}$. We use the slope parameter $r$ to quantify the survival rate of an input initial $v_2^{ini}\{\rm PP\}$. 
We show in Fig.~\ref{fig:3} the survival rate as a function of $p_T$ for $b$=8~fm Au+Au collisions. The survival rate increases with $p_T$. This is because partons with lower $p_T$ suffer on average more collisions before freezeout, which tend to wash out the initial $v_2$. Meanwhile, at higher $p_T$, the survival rate is larger as they suffer fewer collisions. With zero collision, the initial $v_2^{ini}\{\rm PP\}$ will be intact and the survival rate would be 100\%. It is interesting to note that the survival rate at high $p_T$ can be larger than unity, at least for the partons in the highest $p_T$ bin studied in this analysis. This probably suggests some non-linear effect in $v_2$ at high $p_T$ that the initial $v_2$ enhances the final-state development of collective anisotropic flow. It is worthwhile to note that the survival rate of the initial $v_2^{ini}\{\rm PP\}$ to the final-state hadron, at a given $p_T$ value, is smaller than that to the final-state partons before hadronization. This is presumably due to the facts that partons cluster into hadrons at higher $p_T$ and that hadrons rescatter after hadronization.

\begin{figure}[!htb]
  \begin{center}
    \includegraphics[height=4.8cm]{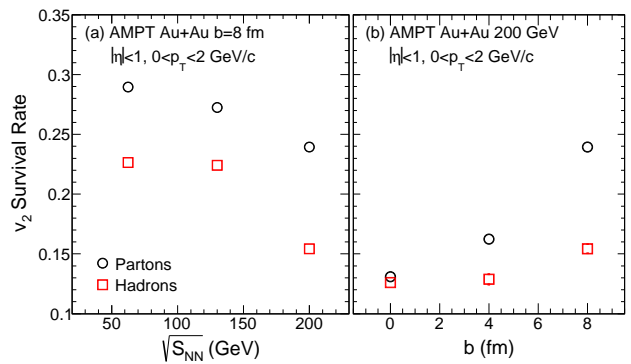}
    \caption{\label{fig:4}(color online) Survival rate as a function of (a) collision energy and (b) impact parameter for mid-rapidity ($|\eta| < 1$) final-state partons and charged hadrons in 0$<p_T<$2 GeV/c for Au+Au collisions by AMPT (string melting).}
  \end{center}
\end{figure}

We show the beam energy dependence of the survival rate in Fig.~\ref{fig:4}~(a) and the centrality dependence in Fig.~\ref{fig:4}~(b). The parton and haron $p_T$ are integrated over the range of 0$<p_T<$2 GeV/c. Note $v_2$ is generally not zero in $b$=0~fm collisions because of event-by-event fluctuations in the initial overlap geometry. The survival rate decreases with increasing collision energy and increasing centrality for both partons and hadrons. Higher energy collisions and/or more central collisions correspond to stronger interactions which reduce the survival rate of the initial anisotropy.

The interpretation of our results is relatively straightforward. The survival rate depends on the final-state activity. The more the partons (and hadrons) interact, the smaller the survival rate of the initial momentum anisotropy. Because in AMPT the collision system has relatively low opacity~\cite{he2015anisotropic}, the survival rate is appreciable. With large opacity, the survival rate should be minimal. It will be interesting to repeat our study using hydrodynamics, starting from a non-zero initial momentum anisotropy.

We have so far used the participant plane from the model to calculate final-state particle $v_2\{\rm PP\}$ to demonstrate that initial momentum anisotropy can survive to the final state. As mentioned earlier, participant plane cannot be accessed in experiment, and instead two-particle correlations are used to measure the momentum anisotropy, $v_2\{2\}$. We have verified that $v_2\{2\}$, while slightly larger than $v_2\{\text{PP}\}$ because of nonflow effects in AMPT, gives the same qualitative conclusion.

In the above we have generated $v_2^{ini}\{\rm PP\}$ with respect to the participant plane. Several physics mechanisms~\cite{Schenke:2015aqa,Fries:2014oya,Ortiz:2013yxa} suggest that the initial momentum anisotropy is independent of the configuration space anisotropy of the collision, but rather random. To study the survival rate of such initial momentum anisotropies, we repeat our analysis but with $v_2^{ini}\{\rm Rnd\}$ with respect to a random plane by Eq.~\ref{eq:6}. 

\begin{figure}[!htb]
  \begin{center}
    \includegraphics[height=4.8cm]{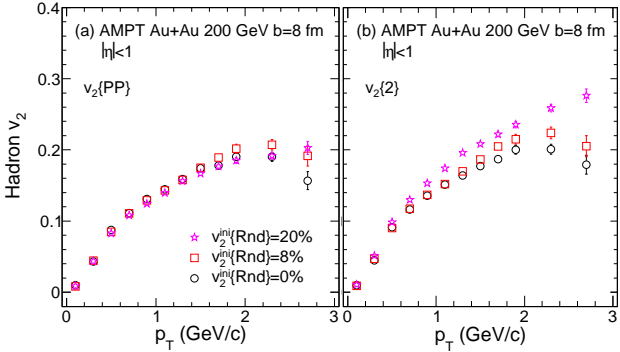}
    \caption{\label{fig:5}(color online) Mid-rapidity ($|\eta| < 1$) hadrons (a) $v_2\{\text{\text{PP}}\}$ and (b) $v_2\{2\}$ versus $p_T$ with $v_2^{ini}\{\text{Rnd}\}$=0\%, 8\% and 20\% for $b$=8~fm Au+Au collisions at $\sqrt{s_{NN}}$=200~GeV by AMPT (string melting).}
  \end{center}
\end{figure}

Figure~\ref{fig:5}~(a) shows the $v_2\{\text{PP}\}$ as a function of $p_T$ with various $v_2^{ini}\{\text{Rnd}\}$. As seen from the figure, $v_2^{ini}\{\text{Rnd}\}$ does not have major effect on $v_2\{\text{PP}\}$. In other words, $v_2^{ini}\{\text{Rnd}\}$ does not survive in $v_2\{\text{PP}\}$. This is understandable because the initial correlations due to $v_2^{ini}\{\text{Rnd}\}$ are averaged out to zero in $v_2\{\text{PP}\}$. Figure~\ref{fig:5}~(b) shows the $v_2\{2\}$ as a function of $p_T$ with various $v_2^{ini}\{\text{Rnd}\}$. It is found that $v_2^{ini}\{\text{Rnd}\}$ has noticeable effect on the final $v_2\{2\}$. 

\begin{figure}[!htb]
  \begin{center}
    \includegraphics[height=5.5cm]{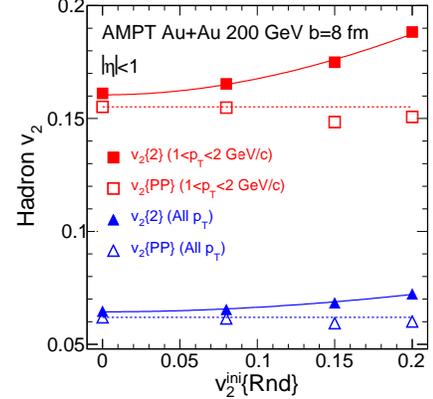}
    \caption{\label{fig:6}(color online) Mid-rapidity ($|\eta| < 1$) hadrons $v_2\{2\}$ and $v_2\{\text{\text{PP}}\}$ versus $v_2^{ini}\{\text{Rnd}\}$ for b=8 fm Au+Au collisions at $\sqrt{s_{NN}}$=200~GeV by AMPT (string melting). The solid lines are fitting results. The dashed lines are horizontal and through the point of $v_2^{ini}\{\text{Rnd}\}$=0.}
  \end{center}
\end{figure}
In order to study the survival rate of $v_2^{ini}\{\text{Rnd}\}$, we show in Fig.~\ref{fig:6} the mid-rapidity ($|\eta|<1$) charged hadron $v_2\{2\}$ (solid markers) and $v_2\{\text{PP}\}$ (hollow markers) versus $v_2^{ini}\{\text{Rnd}\}$ in two $p_T$ ranges. The $v_2\{2\}$ is slightly larger than $v_2\{\text{PP}\}$ with no $v_2^{ini}$, and this is because of nonflow effects in AMPT which are present in $v_2\{2\}$ but not in $v_2\{\text{PP}\}$.  The horizontal dashed line (going through the hollow point at $v_2^{ini}\{\rm Rnd\}$=0) would indicate vanishing effect of $v_2^{ini}\{\text{Rnd}\}$. As can be seen from the figure, the $v_2\{\text{PP}\}$ data points approximately lie along the dashed line. Some minor, but noticeable deviations from the horizontal line may be due to the fact that the AMPT evolution will not be the same once an initial $v_2^{ini}\{\text{Rnd}\}$ is introduced in an event. A large $v_2^{ini}\{\text{Rnd}\}$ may drive the collision system to expand more efficiently in a direction not coinciding with the participant plane but slightly deviating from it. This could cause a smaller $v_2\{\text{PP}\}$, which seems to be the case for large $v_2^{ini}\{\text{Rnd}\}$.

The $v_2\{2\}$, on the other hand, increases with $v_2^{ini}\{\text{Rnd}\}$. The increase is not linear as previously observed with $v_2^{ini}\{\text{PP}\}$. This can be understood as follows. Suppose the final single particle azimuthal distribution is described by
\begin{widetext}
\begin{eqnarray}
dN/d\phi \propto 1+2v_2\{\text{PP}\}(0)\cos(\phi-\Psi_2^{\text{PP}})+2r\times v_2^{ini}\{\text{Rnd}\}\cos2(\phi-\Psi_2^{\text{Rnd}}),
\end{eqnarray}
where $v_2\{\rm PP\}(0)$ is the ``hydrodynamic" flow from normal AMPT, and $r\times v_2^{ini}\{\rm Rnd\}$ is the surviving initial momentum anisotropy along $\Psi_2\{\rm Rnd\}$ with a rate $r$. The two-particle correlations would be
\begin{eqnarray}
d^2N/d\Delta\phi \propto\int d\Psi_2^{\text{PP}}\int d\Psi_2^{\text{Rnd}}  [1+ 2v_2\{\text{PP}\}(0)\cos(\phi_1-\Psi_2^{\text{PP}})+2r\times v_2^{ini}\{\text{Rnd}\}\cos2(\phi_1-\Psi_2^{\text{Rnd}})]\nonumber
\end{eqnarray}
\begin{eqnarray}
\times [1+ 2v_2\{\text{PP}\}(0)\cos(\phi_2-\Psi_2^{\text{PP}})+2r\times v_2^{ini}\{\text{Rnd}\}\cos2(\phi_2-\Psi_2^{\text{Rnd}})]\nonumber
\end{eqnarray}

\begin{eqnarray}
\propto 1+ 2v_2\{\text{PP}\}(0)^2\cos2\Delta\phi+r^2\times 2v_2^{ini}\{\text{Rnd}\}^2\cos2\Delta\phi.
\end{eqnarray}
\end{widetext}
Thus, comparing to Eq.~\ref{eq:999} , we have
\begin{eqnarray}\label{eq:18}
v_2\{2\} =\sqrt{ v_2\{\text{PP}\}(0)^2 + r^2\times v_2^{ini}\{\text{Rnd}\}^2}.
\end{eqnarray}
The relationship is not linear, but quadratic. 
If the initial $v_2^{ini}$ was with respect to the participant plane, as we have studied previously, then the final $v_2$ would be
\begin{eqnarray}\label{eq:19}
v_2\{2\} \approx v_2\{\text{PP}\} = v_2\{\text{PP}\}(0) + r\times v_2^{ini}\{\text{PP}\},
\end{eqnarray}
as we have used in the linear fit in Fig.~\ref{fig:2}. We thus fit the $v_2\{2\}$ vs. $v_2^{ini}\{\text{Rnd}\}$ data in Fig.~\ref{fig:6} by a quadratic function of the form in Eq.~\ref{eq:18}.
The fits are shown by the solid curves in Fig.~\ref{fig:6}. Our results show that $v_2^{ini}\{\text{Rnd}\}$ will survive to final state. This is because the $v_2^{ini}\{\text{Rnd}\}$ introduces an initial particle correlation, which survives to the end when the particles interact minimally in the final state. Therefore, the initial momentum anisotropy, even with respect to a random plane, will contribute to the final anisotropy measurement by particle correlations.



\section{Summary}
We have studied to what extent an input initial momentum anisotropy survives to the final state using the AMPT model. It is found that the final momentum anisotropy shows a linear dependence on the initial momentum anisotropy relative to the configuration-space participant plane (PP). The slope of this linear dependence (survival rate) increases approximately linearly with $p_T$, with~$\sim$100\% survival rate at high $p_T$$\sim$2.5~GeV/c. The survival rate decreases with increasing centrality and increasing beam energy. An initial momentum anisotropy relative to a random plane does not survive to the final-state $v_2\{\text{PP}\}$ with respect to the participant plane, but survives to the two-particle cumulant $v_2\{2\}$. The $v_2\{2\}$ survival rate is quadratic in $v_2^{ini}\{\text{Rnd}\}$ with appreciable magnitude when $v_2^{ini}\{\text{Rnd}\}$ is sizeable. Our results can be understood as that the survival rate decreases with increasing final-state interactions.

\section{Acknowledgments}
The authors acknowledge discussions with Dr.~S.~Schlichting and Dr.~K.~Xiao. This~work~was~supported by the MOST of China under 973 Grant 2015CB856901, the National Natural Science Foundation of China (NSFC) under Grants 11135011, 11228513, 11221504, U.S.~Department of Energy under Grant No.~DE-SC0012910, and CCNU-QLPL Innovation Fund Grant No.~QLPL2015P01.


\end{document}